\documentclass[english]{article}
\usepackage[latin9]{inputenc}
\usepackage{float}
\usepackage{textcomp}
\usepackage{amsmath}
\usepackage{graphicx}
\usepackage{subfig}
\usepackage{stackengine}

\makeatletter

\providecommand{\tabularnewline}{\\}

\usepackage[section]{placeins}

\providecommand{\tabularnewline}{\\}

\usepackage{tikz}
\usepackage{verbatim}
\usetikzlibrary{arrows}
\usepackage{caption}
\usepackage{babel}
\newcommand{\len}{3} 
\newcommand{\spa}{4} 
\newcommand{\off}{0.3} 

\newcommand{\jpi}[2]{#1$^{#2}$}
\newcommand{\iso}[2]{$^{#1}$#2}

\newcommand{\lev}[3]{
\draw [level](0,#1) -- (\len,#1) ;
\draw (\len -3.1,#1) -- (\len -3.3, #1 + #3) -- (\len -3.6, #1 + #3);
\node[left] at (\len -3.6,#1+#3) {#1, #2};
}

\newcommand{\levS}[4]{
\draw [level](#3,#1) -- (#3+\len,#1) ;
\draw (#3+\len +0.1,#1) -- (#3+\len +0.3, #1 + #4) -- (#3+\len +0.6, #1 + #4);
\node[right] at (#3+\len +0.6,#1+#4) {#1, #2};
}

\newcommand{\levL}[4]{
\draw [level](#3,#1) -- (#3+\len,#1) ;
\draw (#3+\len -3.1,#1) -- (#3+\len -3.3, #1 + #4) -- (#3+\len -3.6, #1 + #4);
\node[left] at (#3+\len -3.6,#1+#4) {#1, #2};
}

\providecommand{\tabularnewline}{\\}

\providecommand{\tabularnewline}{\\}

\usepackage{babel}

\makeatother

\usepackage{babel}
\begin{document}

\title{Calculated Inter-band B(E2)'s in the p-f Shell}

\author{Levering Wolfe and Larry Zamick (Corresponding author)\\
 Department of Physics and Astronomy, \\
 Rutgers University, Piscataway, New Jersey 08854}

\maketitle

\section{Abstract}

Whereas previous works for B(E2)'s in the even-even Ti isotopes focused
on yrast transitions we here also consider inter-band transitions
to a second group, i.e. states like 0$_{2}$, 1$_{1}$, 2$_{2}$, 3$_{1}$, 4$_{2}$,
5$_{1}$, etc. We focus on variations from one even-even Ti isotope
to the next. We make a qualitative comparison with similar transitions
in a heavier deformed nucleus.

\section{Introduction}
In the past one of us (LZ) was involved in simple single j shell calculations in the f$_\frac{7}{2}$ shell
e.g. Bayman et al.[1], McCullen et al. [2], see also Sherr et al. [3]. Despite the simplicity of the
model one could find meaningful results for magnetic moments, beta decay transitions [1,2]
and especially spectroscopic factors in single nucleon transfer reactions such as (p,d)[3].
However in the single j shell model ,with j=f$_\frac{7}{2}$, one could not make meaningful calculations
of B(E2) transitions. The rates from the J=0+ ground state of an even -even nucleus to the
first 2$^+$ state were much too weak. It was clear that the collectivity built up in these states
could not be handled in a single j-shell model.
At the other extremes we have the rotational model of Bohr and Mottelson [4]. It is clear from previous work [5] that although there is indeed considerable collectivity,
one has not yet reached the stage where the rotational model is applicable. 

Previous studies of even-even Ti isotopes showed reasonably strong
B(E2)'s in the yrast band: J=0$_{1}$ to 2$_{1}$, 2$_{1}$ to 4$_{1}$,
etc. {[}5{]}. In this work we study transitions from states in the
yrast band to a second group of states: 1$_{1}$, 2$_{2}$, 3$_{1}$, 4$_{2}$, 5$_{1}$, i.e. second excited states of even J and lowest states of odd J. For orientation purposes we show the calculated spectrum of $^{46}$Ti in Fig 1. The shell model code NUSHELLX@MSU
{[}6{]} was used to perform these calculations. Further details are
given in the work of Honma et al. {[}7{]}. We show results for the
GX1A{[}7{]} and FPD6 Interactions{[}8{]}. For both interactions the effective charges of the proton and neutron are 1.5 and 0.5 respectively. Most of the B(E2)'s will be shown with J$_{i}$ less than J$_{f}$. To turn things around one can use the relation

\begin{equation}
(2J_{i}+1) B(E2, J_{i}\rightarrow J_{f}) = (2J_{f}+1)
B(E2, J_{f}\rightarrow J_{i})
\end{equation}

We make comparisons with the rotational model as described by Bohr
and Mottelson{[}4{]}. They give simple formulas for B(E2)'s and static
quadrupole moments (they use I instead of J for angular momentum).

\begin{equation}
B(E2, I_{i}K \rightarrow I_{f} K)= (\frac{5}{16 \pi}) Q_{0}^{2}(I_{i}K
2 0|I_{f}K)^{2}
\end{equation}

\begin{equation}
Q(I,K)= \frac{3K^{2}-I(I+1)}{(I+1) (2I+3)}Q_{0}
\end{equation}

Where Q is the static quadrupole moment Q$_0$ is the intrinsic quadrupole moment. Our first group should be compared with the K=0 band of Bohr and Mottelson although we recognize that the shell model and rotational model are
not exactly the same. Our second group differs from a K=2 band inasmuch
as we include states 0$_{2}$ and odd J's - 1$_{1}$, 3$_{1}$ etc.

Our main concern will be transitions from the yrast band to the second
group for which there are no such clear cut formulas.

The nuclei considered are $^{44}$Ti, $^{46}$Ti, $^{48}$Ti ,$^{50}$Ti
and $^{48}$Cr. Whereas Robinson et al. {[}5{]} considered the even
J yrast band we here show results which also include a second
group as described in the introduction. 

\section{Comments on the two interactions}
The values of B(E2, 0$_1$ $\rightarrow$ 2$_1$) are consistently larger with the FPD6 interaction than
with GX1A. For example in $^{48}$Cr the respective values are 1570 and 1254 e$^2$fm$^4$. A
contributing factor for this can be found by looking at the single particle energies relative
to f$_\frac{7}{2}$ in Table 1. For example the p$_\frac{3}{2}$-f$_\frac{7}{2}$ splitting with FPD6 is 1.8942 MeV, which
is significantly smaller than the corresponding value for GX1A
of 2.9447 MeV. There will therefore be more configuration mixing with FPD6 and this
leads to an enhancement of the B(E2) strength.

\begin{center}

\captionof{table}{Single Particle energies}
\begin{tabular}{|c|c|c|}
\hline
     & FPD6 & GX1A \\
     \hline
    0f$_\frac{7}{2}$ & 0 (-8.3876)&0 (-8.6240) \\
    \hline
    1p$_\frac{3}{2}$ & 1.8942&2.9447 \\
    \hline
    0f$_\frac{5}{2}$&6.4910&7.2411 \\
    \hline
    1p$_\frac{1}{2}$&3.9093&4.4870\\
    \hline
\end{tabular}

\end{center}

\section{The Tables}

In Table 2 we compare B(E2)'s from $J_1$ to $(J+2)_1$ (intra-band) ,
$J_1$ to $(J+2)_2$ (inter-band) and $J_2$ to $(J+2)_2$ (intra-band).
Comparisons are made between the FPD6 and GXIA interactions.

\captionof{table}{Selected B(E2)'s from J to J+2 with the FPD6
interaction (left) and GX1A interaction (right) in e$^2$fm$^4$}

\begin{table}[H]
\begin{minipage}[b]{0.45\linewidth}%
\centering %
\begin{tabular}{|l|l|l|l|}
\hline 
 & $0_1$ to $2_1$  & $0_1$ to $2_2$  & $0_2$ to $2_2$ \tabularnewline
\hline 
$^{44}$Ti  & 699.00  & 0.11  & 212.00 \tabularnewline
\hline 
$^{46}$Ti  & 780.50  & 46.03  & 50.80 \tabularnewline
\hline 
$^{48}$Ti  & 638.00  & 108.40  & 13.50 \tabularnewline
\hline 
$^{50}$Ti  & 569.00  & 0.68  & 0.61 \tabularnewline
\hline 
$^{48}$Cr  & 1570.00  & 52.10  & 0.95\tabularnewline
\hline 
\end{tabular}%
\end{minipage}\hspace{3.25cm} %
\begin{minipage}[b]{0.45\linewidth}%
 \centering %
\begin{tabular}{|l|l|l|}
\hline 
\begin{tabular}{@{}l@{}}
$0_1$ to $2_1$ \tabularnewline
\end{tabular} & $0_1$ to $2_2$  & $0_2$ to $2_2$ \tabularnewline
\hline 
526.00  & 28.30  & 330.00 \tabularnewline
\hline 
624.00  & 4.03  & 161.00 \tabularnewline
\hline 
521.00  & 99.10  & 7.28 \tabularnewline
\hline 
502.00  & 0.07  & 77.90 \tabularnewline
\hline 
1254.00  & 3.08  & 294.00\tabularnewline
\hline 
\end{tabular}%
\end{minipage}
\end{table}

\begin{table}[H]
\begin{minipage}[b]{0.45\linewidth}%
\centering %
\begin{tabular}{|l|l|l|l|}
\hline 
 & $2_1$ to $4_1$  & $2_1$ to $4_2$  & $2_2$ to $4_2$ \tabularnewline
\hline 
$^{44}$Ti  & 343.00  & 1.29  & 118.00 \tabularnewline
\hline 
$^{46}$Ti  & 399.00  & 5.73  & 14.90 \tabularnewline
\hline 
$^{48}$Ti  & 349.00  & 0.36  & 134.00 \tabularnewline
\hline 
$^{50}$Ti  & 212.00  & 0.85  & 18.80 \tabularnewline
\hline 
$^{48}$Cr  & 789.60  & 11.70  & 366.00\tabularnewline
\hline 
\end{tabular}%
\end{minipage}\hspace{3.25cm} %
\begin{minipage}[b]{0.45\linewidth}%
 \centering %
\begin{tabular}{|l|l|l|}
\hline 
\begin{tabular}{@{}l@{}}
$2_1$ to $4_1$\tabularnewline
\end{tabular} & $2_1$ to $4_2$  & $2_2$ to $4_2$ \tabularnewline
\hline 
246.00  & 0.01  & 160.00 \tabularnewline
\hline 
286.00  & 11.70  & 50.60 \tabularnewline
\hline 
269.00  & 5.62  & 83.60 \tabularnewline
\hline 
176.00  & 1.10  & 9.87 \tabularnewline
\hline 
609.00  & 12.60  & 89.20\tabularnewline
\hline 
\end{tabular}%
\end{minipage}
\end{table}

\begin{table}[H]
\begin{minipage}[b]{0.45\linewidth}%
\centering %
\begin{tabular}{|l|l|l|l|}
\hline 
 & $4_1$ to $6_1$  & $4_1$ to $6_2$  & $4_2$ to $6_2$ \tabularnewline
\hline 
$^{44}$Ti  & 232.00  & 9.54  & 87.80 \tabularnewline
\hline 
$^{46}$Ti  & 314.00  & 6.97  & 0.29 \tabularnewline
\hline 
$^{48}$Ti  & 73.03  & 147.00  & 44.12 \tabularnewline
\hline 
$^{50}$Ti  & 88.50  & 15.50  & 107.40 \tabularnewline
\hline 
$^{48}$Cr  & 657.00  & 15.08  & 487.00\tabularnewline
\hline 
\end{tabular}%
\end{minipage}\hspace{3.25cm} %
\begin{minipage}[b]{0.45\linewidth}%
 \centering %
\begin{tabular}{|l|l|l|}
\hline 
\begin{tabular}{@{}l@{}}
$4_1$ to $6_1$\tabularnewline
\end{tabular} & $4_1$ to $6_2$  & $4_2$ to $6_2$ \tabularnewline
\hline 
155.00  & 33.80  & 83.50 \tabularnewline
\hline 
228.00  & 4.38  & 1.54 \tabularnewline
\hline 
87.80  & 95.50  & 57.10 \tabularnewline
\hline 
67.90  & 10.50  & 102.00 \tabularnewline
\hline 
487.00  & 8.91  & 68.90\tabularnewline
\hline 
\end{tabular}%
\end{minipage}
\end{table}

\begin{table}[H]
\begin{minipage}[b]{0.45\linewidth}%
\centering %
\begin{tabular}{|l|l|l|l|}
\hline 
 & $6_1$ to $8_1$  & $6_1$ to $8_2$  & $6_2$ to $8_2$ \tabularnewline
\hline 
$^{44}$Ti  & 146.00  & 1.90  & 47.60 \tabularnewline
\hline 
$^{46}$Ti  & 250.10  & 0.01  & 119.00 \tabularnewline
\hline 
$^{48}$Ti  & 91.50  & 9.06  & 74.50 \tabularnewline
\hline 
$^{50}$Ti  & 0.00  & 31.40  & 5.98 \tabularnewline
\hline 
$^{48}$Cr  & 561.00  & 19.30  & 468.00\tabularnewline
\hline 
\end{tabular}%
\end{minipage}\hspace{3.25cm} %
\begin{minipage}[b]{0.45\linewidth}%
 \centering %
\begin{tabular}{|l|l|l|}
\hline 
\begin{tabular}{@{}l@{}}
$6_1$ to $8_1$ \tabularnewline
\end{tabular} & $6_1$ to $8_2$  & $6_2$ to $8_2$ \tabularnewline
\hline 
94.70  & 2.97  & 39.40 \tabularnewline
\hline 
190.00  & 0.12  & 107.00 \tabularnewline
\hline 
102.00  & 5.70  & 73.90 \tabularnewline
\hline 
0.43  & 14.40  & 2.62 \tabularnewline
\hline 
403.00  & 3.82  & 121.00\tabularnewline
\hline 
\end{tabular}%
\end{minipage}
\end{table}

\begin{table}[H]
\begin{minipage}[b]{0.45\linewidth}%
\centering %
\begin{tabular}{|l|l|l|l|}
\hline 
 & $8_1$ to $10_1$  & $8_1$ to $10_2$  & $8_2$ to $10_2$ \tabularnewline
\hline 
$^{44}$Ti  & 135.00  & 0.64  & 2.87 \tabularnewline
\hline 
$^{46}$Ti  & 180.30  & 2.96  & 68.80 \tabularnewline
\hline 
$^{48}$Ti  & 86.16  & 9.03  & 26.30 \tabularnewline
\hline 
$^{50}$Ti  & 55.90  & 1.16  & 76.70 \tabularnewline
\hline 
$^{48}$Cr  & 434.00  & 13.20  & 1.98\tabularnewline
\hline 
\end{tabular}%
\end{minipage}\hspace{3.25cm} %
\begin{minipage}[b]{0.45\linewidth}%
 \centering %
\begin{tabular}{|l|l|l|}
\hline 
\begin{tabular}{@{}l@{}}
$8_1$ to $10_1$ \tabularnewline
\end{tabular} & $8_1$ to $10_2$  & $8_2$ to $10_2$ \tabularnewline
\hline 
114.00  & 0.14  & 19.50 \tabularnewline
\hline 
134.00  & 15.30  & 79.30 \tabularnewline
\hline 
68.20  & 13.00  & 26.60 \tabularnewline
\hline 
57.50  & 0.27  & 2.53 \tabularnewline
\hline 
261.00  & 20.10  & 16.90\tabularnewline
\hline 
\end{tabular}%
\end{minipage}
\end{table}

\begin{table}[H]
\begin{minipage}[b]{0.45\linewidth}%
\centering %
\begin{tabular}{|l|l|l|l|}
\hline 
 & $10_1$ to $12_1$  & $10_1$ to $12_2$  & $10_2$ to $12_2$ \tabularnewline
\hline 
$^{44}$Ti  & 75.40  & 4.07  & 1.59 \tabularnewline
\hline 
$^{46}$Ti  & 64.40  & 16.20  & 5.49 \tabularnewline
\hline 
$^{48}$Ti  & 34.00  & 14.30  & 0.48 \tabularnewline
\hline 
$^{50}$Ti  & 0.00  & 19.30  & 0.05 \tabularnewline
\hline 
$^{48}$Cr  & 180.10  & 167.00  & 70.70\tabularnewline
\hline 
\end{tabular}%
\end{minipage}\hspace{3.25cm} %
\begin{minipage}[b]{0.45\linewidth}%
 \centering %
\begin{tabular}{|l|l|l|}
\hline 
\begin{tabular}{@{}l@{}}
$10_1$ to $12_1$ \tabularnewline
\end{tabular} & $10_1$ to $12_2$  & $10_2$ to $12_2$ \tabularnewline
\hline 
64.10  & 1.87  & 2.09 \tabularnewline
\hline 
49.84  & 1.96  & 13.90 \tabularnewline
\hline 
28.80  & 1.29  & 0.15 \tabularnewline
\hline 
56.50  & 15.50  & 2.14 \tabularnewline
\hline 
194.00  & 40.20  & 62.80\tabularnewline
\hline 
\end{tabular}%
\end{minipage}
\end{table}

\begin{table}[H]
\begin{minipage}[b]{0.45\linewidth}%
\centering %
\begin{tabular}{|l|l|l|l|}
\hline 
 & $12_1$ to $14_1$  & $12_1$ to $14_2$  & $12_2$ to $14_2$ \tabularnewline
\hline 
$^{46}$Ti  & 45.70  & 4.44  & 8.28 \tabularnewline
\hline 
$^{48}$Ti  & 5.49  & 4.49  & 0.08 \tabularnewline
\hline 
$^{50}$Ti  & 0.04  & 0.64  & 2.81 \tabularnewline
\hline 
$^{48}$Cr  & 160.40  & 12.60  & 182.00\tabularnewline
\hline 
\end{tabular}%
\end{minipage}\hspace{3.25cm} %
\begin{minipage}[b]{0.45\linewidth}%
\begin{tabular}{|l|l|l|}
\hline 
\begin{tabular}{@{}l@{}}
$12_1$ to $14_1$ \tabularnewline
\end{tabular} & $12_1$ to $14_2$  & $12_2$ to $14_2$ \tabularnewline
\hline 
44.95  & 1.96  & 0.11 \tabularnewline
\hline 
5.26  & 0.04  & 0.48 \tabularnewline
\hline 
13.60  & 0.86  & 8.09 \tabularnewline
\hline 
148.00  & 3.54  & 50.70\tabularnewline
\hline 
\end{tabular}%
\end{minipage}
\end{table}

\begin{table}[H]
\begin{minipage}[b]{0.45\linewidth}%
\centering %
\begin{tabular}{|l|l|l|l|}
\hline 
 & $14_1$ to $16_1$  & $14_1$ to $16_2$  & $14_2$ to $16_2$ \tabularnewline
\hline 
$^{46}$Ti  & 1.30  & 0.05  & 0.32 \tabularnewline
\hline 
$^{48}$Ti  & 49.50  & 3.05  & 0.01 \tabularnewline
\hline 
$^{50}$Ti  & 58.07  & 0.04  & 0.03 \tabularnewline
\hline 
$^{48}$Cr  & 78.60  & 12.40  & 168.00\tabularnewline
\hline 
\end{tabular}%
\end{minipage}\hspace{3.25cm} %
\begin{minipage}[b]{0.45\linewidth}%
\begin{tabular}{|l|l|l|}
\hline 
\begin{tabular}{@{}l@{}}
$14_1$ to $16_1$ \tabularnewline
\end{tabular} & $14_1$ to $16_2$  & $14_2$ to $16_2$ \tabularnewline
\hline 
0.62  & 0.00  & 0.86 \tabularnewline
\hline 
7.25  & 2.95  & 0.26 \tabularnewline
\hline 
21.97  & 0.06  & 11.50 \tabularnewline
\hline 
71.30  & 8.14  & 112.00\tabularnewline
\hline 
\end{tabular}%
\end{minipage}
\end{table}

\section{Discussion of the B(E2) Tables}

We start with a crude overview of the results. For the yrast transitions
if we go in the opposite direction, namely from J to (J-2), then in
the rotational and vibrational models the B(E2)'s increase with J but
in our shell model calculations they decrease with J after J=4 {[}4{]}.

On the whole the $J_1$ to $(J+2)_1$ (yrast) transitions are the largest
and the $J_2$ to $(J+2)_2$ are often large as well but not as much.
To partially understand this we use the rotational formula for B(E2)'s
as given in the introduction.

Consider for example a transition from J=2 to J=4. If we assume Band 1 has K=0 and band 2 has K=2, and both have the same intrinsic quadrupole moment Q$_{0}$, then the ratio

\begin{equation}
    \frac{B(E2,2_2\rightarrow4_2)}{ B(E2,2_1\rightarrow4_1)}= [{\frac{(2\hspace{3 pt}2\hspace{3 pt} 2\hspace{3 pt} 0| 4\hspace{3 pt}2)}{(2\hspace{3 pt} 0\hspace{3 pt} 2\hspace{3 pt} 0|4\hspace{3 pt}0)}}]^{2}=\frac{5}{12}=0.41667.
\end{equation}

In detail from Table 2, the ratios are smaller than that except for
$^{44}$ Ti and $^{48}$Cr.

Although some of the $J_2$ to $(J+2)_2$ transitions are reasonably large this is not always the case. For
example consider the $2_2$ to $4_2$ transitions using the FPD6 interaction.
The values for $^{44}$Ti, $^{46}$Ti, $^{48}$Ti and $^{50}$Ti, and $^{48}$Cr are respectively, 118.0,
14.90, 134.0, 18.8, and 360.0 e$^{2}$fm$^{4}$. Some are large and
some are small. 

One main point of this study is that the inter-band transitions are
quite small with some glaring exceptions e.g. $0_1$ to $2_2$ in $^{48}$Ti
-- 108.40 with FPD6 and 99.1 e$^{2}$ fm$^{4}$with GX1A. It is difficult
to see a simple trend with neutron number for these inter-band transitions.
Using again the example of $2_1$ to $4_2$ the respective values for
FPD6 are 1.29, 5.73, 0.36, 0.85 and 11.70 e$^{2}$ fm$^{4}$. In the Ti
isotopes we go from low to high to low to high, so it is difficult to find a
clear-cut trend. The results may be useful however to prevent excessive
hand waving.

Also by using 2 interactions we get a feel about how far we can go
in making quantitative remarks about the inter-band transitions. There
is unfortunately much variation in the results. For example, again for $2_1$ to $4_2$, the values for FPD6 (GX1A) are respectively: 1.90 (0.01), 5.73 (11.70), 0.36 (5.62), 0.85 (1.10), 11.70 (12.60).
We can however make the qualitative remarks that the $J_1$ to $(J+2)_2$ B(E2)'s are much smaller than the yrast B(E2)'s and in the majority of cases also smaller than the $J_2$ to $(J+2)_2$ B(E2)'s.

\section{Comparisons with Collective Models}

The shell model calculations can be compared with predictions  of the rotational model.  The rotational expression is as follows:

$$\frac{B(E2, J \rightarrow (J+2))}{B(E2, 0 \rightarrow 2)} = 3 \frac{(J+2)(J+1)^2}{(2J+1)(2J+2)(2J+3)} $$

The values from J=0 to 16 are respectively

\begin{table}[h]
\captionof{table}{Ratio of E2 transitions from J to (J+2) over E2 transition from 0 to 2}
\centering
\begin{tabular}{|l|l|}
\hline
J  & $\frac{B(E2, J \rightarrow (J+2))}{B(E2, 0 \rightarrow 2)}$        \\ \hline
0  & 1.000 \\ \hline
2  & 0.514 \\ \hline
4  & 0.455 \\ \hline
6  & 0.431 \\ \hline
8  & 0.418 \\ \hline
10 & 0.410 \\ \hline
12 & 0.404 \\ \hline
14 & 0.400 \\ \hline
16 & 0.397 \\ \hline
\end{tabular}
\end{table}
\flushleft

For the transition from J=14 to 16 the value is 0.400. On the other hand the value of this
ratio for say $^{48}$Cr with FPD6 is $\frac{78.6}{1570}$=0.050. There is a tremendous difference. 
\\
This discrepancy is also present in the vibrational model, even more so. $\frac{B(E2, J+2 \rightarrow J)}{B(E2,2 \rightarrow 0)} = \frac{J+2}{J}$. This is equivalent  to saying  that  B(E2,J+2 $\rightarrow$ J) is proportional to the number of phonons that are present in the initial state. We see that although comparisons with collective models are suggestive there are important differences, and shell model calculations of increasing complexity are needed to get quantitative results in this part of the periodic table.

\section{Comparisons with other work}

In Table 4 we make a comparison of the behavior in the Ti isotopes with
what occurs in more deformed nuclei. It is convenient to choose the
work of Cl\'ement et al. {[}9{]} on $^{98}$Sr because they show several
measured B(E2)'s between states in the yrast band and those in the
next band. The comparison is somewhat hybrid because we are listing
experimental results for Sr and theoretical results for Ti. The B(E2's)
in Weisskopf units (WU) are 19.4 in $^{46}$Ti and 95.5 in $^{98}$Sr.
This shows that the latter nucleus is indeed more strongly deformed
than any of the Ti isotopes.

In their Table 4 Cl\'ement et al. {[}9{]} show reduced matrix elements.
In our Table 3 we show rather the ratio of a given B(E2) to the intra-band
0$_1 \longmapsto$ 2$_1$ B(E2) with the GX1A interaction. The ratio of this transition to $2_1\longmapsto2_2$
in $^{98}$Sr is quite small whereas for $^{44}$Ti and $^{46}$Ti
the values are 0.2909 and 0.1694 respectively. A Ratio close to 0.2
is also found for $0_1\rightarrow2_2$ in$^{48}$Ti.

\begin{center}
\captionof{table}{Ratio $\frac{B(E2)}{B(E2, 0_1\rightarrow2_1)}$} %
\begin{tabular}{|c|c|c|c|c|c|c|}
\hline 
Ji$\rightarrow$Jf  & $^{98}$Sr  & $^{44}$Ti  & $^{46}$Ti  & $^{48}$Ti  & $^{50}$Ti  & $^{48}$Cr\tabularnewline
\hline 
0$_1$$\rightarrow$2$_2$  & 0.00799  & 0.05380  & 0.006458  & 0.1902  & 0.000142  & 0.00246\tabularnewline
\hline 
2$_1$$\rightarrow$0$_2$  & 0.02556  & 0.0113  & 0.0208  & 0.0195  & 0.00219  & 0.0221\tabularnewline
\hline 
2$_1$$\rightarrow$2$_2$  & 0.000767  & 0.2909  & 0.1694  & 0.0845  & 0.00703  & 0.0451\tabularnewline
\hline 
4$_1$$\rightarrow$2$_2$  & 0.004603  & 0.02567  & 0.01651  & 0.00263  & 0.00365  & 0.00606\tabularnewline
\hline 
2$_1$$\rightarrow$3$_1$  &  & 0.0123  & 0.009295  & 0.0595  & 0.000448  & 0.0151\tabularnewline
\hline 
4$_1$$\rightarrow$3$_1$  &  & 0.04297  & 0.006490  & 0.04626  & 0.0307  & 0.00177\tabularnewline
\hline 
\end{tabular}
\par\end{center}

\section{Even J to Odd J Transitions}

In Table 5 we show some selected even J to odd J B(E2)'s. We show in Table
5 some selected ones in $^{48}$Cr. While most of them are small,
there are some surprisingly large ones from $4_2$ to $5_1$, $6_2$ to
$5_1$, and $6_2$ to $7_1$.

\begin{center}
\captionof{table}{B(E2)'s from even J to odd J in $^{48}$Cr in
e$^{2}$fm$^{4}$} %
\begin{tabular}{|c|c|c|c|}
\hline 
J$_{i}$  & J$_{f}$  & GX1A B(E2) & FPD6 B(E2)\tabularnewline
\hline 
2$_1$  & 1$_1$  & 11.4 & 4.8\tabularnewline
\hline 
2$_1$  & 3$_1$  & 19.0 & 21.8\tabularnewline
\hline 
4$_1$  & 3$_1$  & 2.22 & 7.18\tabularnewline
\hline 
4$_1$  & 5$_1$  & 11.8 & 16.4\tabularnewline
\hline 
6$_1$  & 5$_1$  & 15.6 & 9.1\tabularnewline
\hline 
6$_1$  & 7$_1$  & 1.29 & 12.7\tabularnewline
\hline 
2$_2$  & 1$_1$  & 0.0003 & 12.3\tabularnewline
\hline 
2$_2$  & 3$_1$  & 11.2 & 840\tabularnewline
\hline 
4$_2$  & 3$_1$  & 2.39 & 427\tabularnewline
\hline 
4$_2$  & 5$_1$  & 206 & 338\tabularnewline
\hline 
6$_2$  & 5$_1$  & 150 & 199\tabularnewline
\hline 
6$_2$  & 7$_1$  & 182 & 138\tabularnewline
\hline 
\end{tabular}
\par\end{center}

\section{Electric Quadrupole Moments}

\begin{table}[H]
\begin{minipage}[b]{0.45\linewidth}%
\captionof{table}{FPD6 quadrupole moments in e fm$^2$} %
\begin{tabular}{|l|l|l|l|l|l|}
\hline 
 & $^{44}$Ti  & $^{46}$Ti  & $^{48}$Ti  & $^{50}$Ti  & $^{48}$Cr \tabularnewline
\hline 
$2_1$  & -21.60  & -23.60  & -18.90  & 3.64  & -35.50 \tabularnewline
\hline 
$2_2$  & 14.40  & 3.71  & 2.34  & 12.90  & 36.98 \tabularnewline
\hline 
\end{tabular}%
\end{minipage}\hspace{2.0cm} %
\begin{minipage}[b]{0.45\linewidth}%
\centering \captionof{table}{GX1A quadrupole moments e fm$^2$} %
\begin{tabular}{|l|l|l|l|l|l|}
\hline 
 & $^{44}$Ti  & $^{46}$Ti  & $^{48}$Ti  & $^{50}$Ti  & $^{48}$Cr \tabularnewline
\hline 
$2_1$  & -6.01  & -13.6  & -14.5  & 6.53  & -30.8 \tabularnewline
\hline 
$2_2$  & -0.89  & 7.1  & 5.02  & 13.3  & 21.9 \tabularnewline
\hline 
\end{tabular}%
\end{minipage}
\end{table}

In Tables 6 and 7 we show results for static quadrupole moments of 2$_1$ and 2$_2$ states. Note that for $^{46}$Ti,$^{48}$Ti and $^{48}$Cr the quadrupole
moments of the $2_1$ states are negative and those of the $2_2$ states
are positive. In $^{44}$Ti with FPD6 we get the same behavior as in other nuclei, Q$_1$ negative and Q$_2$ positive, but this is not the case with GX1A. We get negative values for both Q's with GX1A. This may be explained in part by the fact that with only 4  valence nucleons it is difficult to establish collective motion and, as mentioned before, the single particle energy splittings are larger with GX1A than with FPD6. This favors single particle behavior over collective behavior for GX1A.
\\
In the rotational model (see introduction) the value of Q(2$_1$)is equal to $-\frac{2}{7}$ Q$_{0}$ whilst the value for Q(2$_2$) is +$\frac{2}{7}$ Q$_{0}$. Indeed, the quadrupole momentof J=2$^+$ for a K=2 band are equal and opposite of those of a K=0 band. Note that for the K=0 band the laboratory quadruple moment is of opposite sign to the intrinsic quadrupole moment. One can
imagine a pencil which clearly has a positive quadruple moment along it's axis.
(prolate). But if one rotates the pencil about an axis perpendicular to the line of the
pencil and about its center of mass one traces out a pancake which will have a negative
quadrupole moment.

\section{B(E2)'s from the J=0+ ground state to several 2$^+$ states}

In Tables 8 to 17 we present B(E2)'s, Energies and B(E2)*Energy for the J=0+ ground
state to 15 J=2$^+$ excited states. In all cases the largest transition is to the 2$_1$ state. After that the 2 interactions
sometimes differ in which state has the next strongest strength. For example in $^{48}$Cr the
first 3 B(E2)'s with FPD6 are 1569, 52.10 and 15.50 e$^2$ fm$^4$ whereas with GX1A they
are 1254, 3.10 and 75.60 e$^2$ fm$^4$. With FPD6 the second 2$^+$ state has the second
most strength but with GX1A it is the third. Things are steadier if we look at the summed
strength and summed energy weighted strength.

\begin{table}[H]
\begin{minipage}[b]{0.45\linewidth}%
\captionof{table}{B(E2) and Energy Weighted B(E2) of transitions from 0$_1$ in $^{44}$Ti with FPD6 interaction in
e$^{2}$fm$^{4}$ MeV}%
\begin{tabular}{|l|l|l|l|}
\hline
        & B(E2)    & Energy & B(E2) * Energy \\ \hline
2$_{1}$    & 698.90   & 1.30   & 908.43         \\ \hline
2$_{2}$    & 0.11     & 4.34   & 0.47           \\ \hline
2$_{3}$    & 10.28    & 6.11   & 62.78          \\ \hline
2$_{4}$    & 4.24     & 7.07   & 29.92          \\ \hline
2$_{5}$    & 6.36     & 7.99   & 50.83          \\ \hline
2$_{6}$    & 6.22     & 8.11   & 50.48          \\ \hline
2$_{7}$    & 1.51     & 8.34   & 12.60          \\ \hline
2$_{8}$    & 11.20    & 9.29   & 104.09         \\ \hline
2$_{9}$    & 0.14     & 9.75   & 1.37           \\ \hline
2$_{10}$   & 0.01     & 9.92   & 0.11           \\ \hline
2$_{11}$   & 0.02     & 10.02  & 0.23           \\ \hline
2$_{12}$   & 0.13     & 10.17  & 1.29           \\ \hline
2$_{13}$   & 0.07     & 10.44  & 0.74           \\ \hline
2$_{14}$   & 0.00     & 10.48  & 0.00           \\ \hline
2$_{15}$   & 0.34     & 10.69  & 3.59           \\ \hline
SUM(15) & 739.53   & x      & 1226.93        \\ \hline
SUM(50) & 747.2302 & x      & 1318.23        \\ \hline
\end{tabular}%
\end{minipage}\hspace{2.5cm} %
\begin{minipage}[b]{0.45\linewidth}%
\captionof{table}{B(E2) and Energy Weighted B(E2) of transitions from 0$_1$ in $^{44}$Ti with GX1A interaction in e$^{2}$fm$^{4}$ MeV}%
\begin{tabular}{|l|l|l|l|}
\hline
        & B(E2)    & Energy & B(E2) * Energy \\ \hline
2$_{1}$    & 526.20   & 1.29   & 677.43         \\ \hline
2$_{2}$    & 28.34    & 3.17   & 89.78          \\ \hline
2$_{3}$    & 14.82    & 5.30   & 78.51          \\ \hline
2$_{4}$    & 3.13     & 6.49   & 20.34          \\ \hline
2$_{5}$    & 0.88     & 6.77   & 5.92           \\ \hline
2$_{6}$    & 11.74    & 7.24   & 84.99          \\ \hline
2$_{7}$    & 0.06     & 7.89   & 0.51           \\ \hline
2$_{8}$    & 0.01     & 8.55   & 0.09           \\ \hline
2$_{9}$    & 0.57     & 8.87   & 5.01           \\ \hline
2$_{10}$   & 14.22    & 8.95   & 127.23         \\ \hline
2$_{11}$   & 0.11     & 9.35   & 1.00           \\ \hline
2$_{12}$   & 4.40     & 9.46   & 41.63          \\ \hline
2$_{13}$   & 0.77     & 9.47   & 7.30           \\ \hline
2$_{14}$   & 4.04     & 9.66   & 39.01          \\ \hline
2$_{15}$   & 0.01     & 9.74   & 0.07           \\ \hline
SUM(15) & 609.29   & x      & 1178.82        \\ \hline
SUM(50) & 617.3805 & x      & 1271.913       \\ \hline
\end{tabular}%
\end{minipage}
\end{table}

\begin{table}[H]
\begin{minipage}[b]{0.45\linewidth}%
\captionof{table}{B(E2) and Energy Weighted B(E2) of transitions from 0$_1$ in $^{46}$Ti with FPD6 interaction in
e$^{2}$fm$^{4}$ MeV}%
\begin{tabular}{|l|l|l|l|}
\hline
        & B(E2)    & Energy & B(E2) * Energy \\ \hline
2$_{1}$    & 780.50   & 0.98   & 762.24         \\ \hline
2$_{2}$    & 46.03    & 3.23   & 148.59         \\ \hline
2$_{3}$    & 1.51     & 3.89   & 5.87           \\ \hline
2$_{4}$    & 7.56     & 4.34   & 32.83          \\ \hline
2$_{5}$    & 0.38     & 4.91   & 1.86           \\ \hline
2$_{6}$    & 0.05     & 5.53   & 0.27           \\ \hline
2$_{7}$    & 3.75     & 5.92   & 22.21          \\ \hline
2$_{8}$    & 7.24     & 6.22   & 45.00          \\ \hline
2$_{9}$    & 6.95     & 6.34   & 44.09          \\ \hline
2$_{10}$   & 10.34    & 6.62   & 68.44          \\ \hline
2$_{11}$   & 2.37     & 6.85   & 16.26          \\ \hline
2$_{12}$   & 0.22     & 7.00   & 1.52           \\ \hline
2$_{13}$   & 0.12     & 7.48   & 0.87           \\ \hline
2$_{14}$   & 0.27     & 7.69   & 2.07           \\ \hline
2$_{15}$   & 0.03     & 7.93   & 0.25           \\ \hline
SUM(15) & 867.31   & x      & 1152.35        \\ \hline
SUM(50) & 875.9495 & x      & 1234.837       \\ \hline
\end{tabular}%
\end{minipage}\hspace{2.5cm} %
\begin{minipage}[b]{0.45\linewidth}%
\captionof{table}{B(E2) and Energy Weighted B(E2) of transitions from 0$_1$ in $^{46}$Ti with GX1A interaction in e$^{2}$fm$^{4}$ MeV}%
\begin{tabular}{|l|l|l|l|}
\hline
        & B(E2)    & Energy & B(E2) * Energy \\ \hline
2$_{1}$    & 624.40   & 1.01   & 627.81         \\ \hline
2$_{2}$    & 4.03     & 2.59   & 10.42          \\ \hline
2$_{3}$    & 43.56    & 3.39   & 147.48         \\ \hline
2$_{4}$    & 8.22     & 4.28   & 35.17          \\ \hline
2$_{5}$    & 0.04     & 5.01   & 0.19           \\ \hline
2$_{6}$    & 21.26    & 5.44   & 115.73         \\ \hline
2$_{7}$    & 0.29     & 5.54   & 1.58           \\ \hline
2$_{8}$    & 0.36     & 5.82   & 2.09           \\ \hline
2$_{9}$    & 1.74     & 6.10   & 10.63          \\ \hline
2$_{10}$   & 5.44     & 6.38   & 34.70          \\ \hline
2$_{11}$   & 6.44     & 6.53   & 42.05          \\ \hline
2$_{12}$   & 0.70     & 6.65   & 4.66           \\ \hline
2$_{13}$   & 1.03     & 6.90   & 7.13           \\ \hline
2$_{14}$   & 0.59     & 7.03   & 4.17           \\ \hline
2$_{15}$   & 0.11     & 7.13   & 0.76           \\ \hline
SUM(15) & 718.20   & x      & 1044.57        \\ \hline
SUM(50) & 733.6302 & x      & 1176.939       \\ \hline
\end{tabular}%
\end{minipage}
\end{table}

\begin{table}[H]
\begin{minipage}[b]{0.45\linewidth}%
\captionof{table}{B(E2) and Energy Weighted B(E2) of transitions from 0$_1$ in $^{48}$Ti with FPD6 interaction in
e$^{2}$fm$^{4}$ MeV}%
\begin{tabular}{|l|l|l|l|}
\hline
        & B(E2)    & Energy & B(E2) * Energy \\ \hline
2$_{1}$    & 638.30   & 1.18   & 751.15         \\ \hline
2$_{2}$    & 108.40   & 2.47   & 267.27         \\ \hline
2$_{3}$    & 10.86    & 3.72   & 40.39          \\ \hline
2$_{4}$    & 3.91     & 4.16   & 16.25          \\ \hline
2$_{5}$    & 2.85     & 4.87   & 13.90          \\ \hline
2$_{6}$    & 20.12    & 5.21   & 104.87         \\ \hline
2$_{7}$    & 4.33     & 5.65   & 24.48          \\ \hline
2$_{8}$    & 0.14     & 5.88   & 0.82           \\ \hline
2$_{9}$    & 1.80     & 6.04   & 10.87          \\ \hline
2$_{10}$   & 1.27     & 6.13   & 7.78           \\ \hline
2$_{11}$   & 1.06     & 6.32   & 6.71           \\ \hline
2$_{12}$   & 0.01     & 6.55   & 0.10           \\ \hline
2$_{13}$   & 0.03     & 7.00   & 0.21           \\ \hline
2$_{14}$   & 0.67     & 7.03   & 4.68           \\ \hline
2$_{15}$   & 1.59     & 7.21   & 11.44          \\ \hline
SUM(15) & 795.34   & x      & 1260.92        \\ \hline
SUM(50) & 811.7956 & x      & 1392.518       \\ \hline
\end{tabular}%
\end{minipage}\hspace{2.5cm} %
\begin{minipage}[b]{0.45\linewidth}%
\captionof{table}{B(E2) and Energy Weighted B(E2) of transitions from 0$_1$ in $^{48}$Ti with GX1A interaction in e$^{2}$fm$^{4}$ MeV}%
\begin{tabular}{|l|l|l|l|}
\hline
        & B(E2)    & Energy & B(E2) * Energy \\ \hline
2$_{1}$    & 520.80   & 1.01   & 525.90         \\ \hline
2$_{2}$    & 99.13    & 2.18   & 216.11         \\ \hline
2$_{3}$    & 26.51    & 3.32   & 87.94          \\ \hline
2$_{4}$    & 0.13     & 4.03   & 0.52           \\ \hline
2$_{5}$    & 19.56    & 4.52   & 88.42          \\ \hline
2$_{6}$    & 24.18    & 4.71   & 113.87         \\ \hline
2$_{7}$    & 0.70     & 5.21   & 3.64           \\ \hline
2$_{8}$    & 0.77     & 5.67   & 4.37           \\ \hline
2$_{9}$    & 3.38     & 5.78   & 19.51          \\ \hline
2$_{10}$   & 0.35     & 5.93   & 2.10           \\ \hline
2$_{11}$   & 0.59     & 6.14   & 3.62           \\ \hline
2$_{12}$   & 0.69     & 6.23   & 4.32           \\ \hline
2$_{13}$   & 0.22     & 6.59   & 1.43           \\ \hline
2$_{14}$   & 1.73     & 6.69   & 11.55          \\ \hline
2$_{15}$   & 1.21     & 6.70   & 8.11           \\ \hline
SUM(15) & 699.94   & x      & 1091.40        \\ \hline
SUM(50) & 721.5041 & x      & 1259.665       \\ \hline
\end{tabular}%
\end{minipage}
\end{table}

\begin{table}[H]
\begin{minipage}[b]{0.45\linewidth}%
\captionof{table}{B(E2) and Energy Weighted B(E2) of transitions from 0$_1$ in $^{50}$Ti with FPD6 interaction in
e$^{2}$fm$^{4}$ MeV}%
\begin{tabular}{|l|l|l|l|}
\hline
        & B(E2)    & Energy & B(E2) * Energy \\ \hline
2$_{1}$    & 568.80   & 1.83   & 1039.65        \\ \hline
2$_{2}$    & 0.68     & 4.19   & 2.84           \\ \hline
2$_{3}$    & 57.08    & 4.56   & 260.18         \\ \hline
2$_{4}$    & 9.13     & 5.08   & 46.40          \\ \hline
2$_{5}$    & 1.32     & 5.92   & 7.79           \\ \hline
2$_{6}$    & 0.35     & 6.23   & 2.16           \\ \hline
2$_{7}$    & 2.90     & 6.46   & 18.74          \\ \hline
2$_{8}$    & 0.00     & 6.63   & 0.02           \\ \hline
2$_{9}$    & 0.63     & 6.78   & 4.29           \\ \hline
2$_{10}$   & 2.63     & 6.97   & 18.30          \\ \hline
2$_{11}$   & 0.01     & 6.97   & 0.06           \\ \hline
2$_{12}$   & 3.05     & 7.16   & 21.86          \\ \hline
2$_{13}$   & 3.03     & 7.43   & 22.53          \\ \hline
2$_{14}$   & 4.66     & 7.55   & 35.16          \\ \hline
2$_{15}$   & 0.05     & 7.67   & 0.35           \\ \hline
SUM(15) & 654.30   & x      & 1480.32        \\ \hline
SUM(50) & 662.7534 & x      & 1556.158       \\ \hline
\end{tabular}%
\end{minipage}\hspace{2.5cm} %
\begin{minipage}[b]{0.45\linewidth}%
\captionof{table}{B(E2) and Energy Weighted B(E2) of transitions from 0$_1$ in $^{50}$Ti with GX1A interaction in e$^{2}$fm$^{4}$ MeV}%
\begin{tabular}{|l|l|l|l|}
\hline
        & B(E2)    & Energy & B(E2) * Energy \\ \hline
2$_{1}$    & 502.30   & 1.62   & 815.74         \\ \hline
2$_{2}$    & 0.08     & 3.90   & 0.30           \\ \hline
2$_{3}$    & 56.92    & 4.22   & 240.07         \\ \hline
2$_{4}$    & 2.53     & 5.03   & 12.71          \\ \hline
2$_{5}$    & 6.38     & 5.31   & 33.86          \\ \hline
2$_{6}$    & 1.23     & 5.88   & 7.25           \\ \hline
2$_{7}$    & 1.78     & 6.04   & 10.71          \\ \hline
2$_{8}$    & 1.57     & 6.53   & 10.27          \\ \hline
2$_{9}$    & 1.60     & 6.60   & 10.55          \\ \hline
2$_{10}$   & 6.41     & 6.74   & 43.26          \\ \hline
2$_{11}$   & 0.09     & 6.80   & 0.63           \\ \hline
2$_{12}$   & 2.98     & 6.93   & 20.67          \\ \hline
2$_{13}$   & 3.35     & 7.07   & 23.72          \\ \hline
2$_{14}$   & 0.64     & 7.20   & 4.60           \\ \hline
2$_{15}$   & 0.01     & 7.37   & 0.09           \\ \hline
SUM(15) & 587.88   & x      & 1234.43        \\ \hline
SUM(50) & 604.1513 & x      & 1373.675       \\ \hline
\end{tabular}%
\end{minipage}
\end{table}

\begin{table}[H]
\begin{minipage}[b]{0.45\linewidth}%
\captionof{table}{B(E2) and Energy Weighted B(E2) of transitions from 0$_1$ in $^{48}$Cr with FPD6 interaction in
e$^{2}$fm$^{4}$ MeV}%
\begin{tabular}{|l|l|l|l|}
\hline
        & B(E2)   & Energy & B(E2) * Energy \\ \hline
2$_{1}$    & 1569.00 & 0.79   & 1237.78        \\ \hline
2$_{2}$    & 52.11   & 3.66   & 190.97         \\ \hline
2$_{3}$    & 14.96   & 4.57   & 68.32          \\ \hline
2$_{4}$    & 17.00   & 5.55   & 94.42          \\ \hline
2$_{5}$    & 12.43   & 6.35   & 78.96          \\ \hline
2$_{6}$    & 1.21    & 6.67   & 8.08           \\ \hline
2$_{7}$    & 20.35   & 6.96   & 141.69         \\ \hline
2$_{8}$    & 0.03    & 7.40   & 0.21           \\ \hline
2$_{9}$    & 1.95    & 7.47   & 14.60          \\ \hline
2$_{10}$   & 0.00    & 7.54   & 0.00           \\ \hline
2$_{11}$   & 12.20   & 7.69   & 93.87          \\ \hline
2$_{12}$   & 0.04    & 7.90   & 0.30           \\ \hline
2$_{13}$   & 5.47    & 8.06   & 44.12          \\ \hline
2$_{14}$   & 2.35    & 8.13   & 19.14          \\ \hline
2$_{15}$   & 0.99    & 8.33   & 8.26           \\ \hline
SUM(15) & 1710.10 & x      & 2000.70        \\ \hline
SUM(50) & 1737.61 & x      & 2266.909       \\ \hline
\end{tabular}%
\end{minipage}\hspace{2.5cm} %
\begin{minipage}[b]{0.45\linewidth}%
\captionof{table}{B(E2) and Energy Weighted B(E2) of transitions from 0$_1$ in $^{48}$Cr with GX1A interaction in e$^{2}$fm$^{4}$ MeV}%
\begin{tabular}{|l|l|l|l|}
\hline
        & B(E2)    & Energy & B(E2) * Energy \\ \hline
2$_{1}$    & 1254.00  & 0.79   & 989.16         \\ \hline
2$_{2}$    & 3.09     & 3.39   & 10.50          \\ \hline
2$_{3}$    & 75.59    & 4.10   & 309.81         \\ \hline
2$_{4}$    & 34.57    & 4.62   & 159.86         \\ \hline
2$_{5}$    & 12.04    & 5.50   & 66.17          \\ \hline
2$_{6}$    & 9.39     & 5.69   & 53.42          \\ \hline
2$_{7}$    & 1.44     & 5.98   & 8.63           \\ \hline
2$_{8}$    & 0.21     & 6.41   & 1.37           \\ \hline
2$_{9}$    & 1.35     & 6.77   & 9.13           \\ \hline
2$_{10}$   & 0.02     & 6.78   & 0.15           \\ \hline
2$_{11}$   & 2.05     & 6.90   & 14.12          \\ \hline
2$_{12}$   & 9.81     & 6.98   & 68.45          \\ \hline
2$_{13}$   & 6.73     & 7.14   & 48.03          \\ \hline
2$_{14}$   & 0.07     & 7.21   & 0.49           \\ \hline
2$_{15}$   & 0.29     & 7.41   & 2.18           \\ \hline
SUM(15) & 1410.66  & x      & 1741.47        \\ \hline
SUM(50) & 1446.022 & x      & 2041.868       \\ \hline
\end{tabular}%
\end{minipage}
\end{table}

We now comment on the tables in this section-B(E2)'s from the J=0+ ground state to many 2$^+$ excited states. Note that with both interactions the overwhelming fraction of the strength is to the first 2$^+$ state. For example in $^{46}$Ti with FPD6 the B(E2) to the first 2$^+$ state is 780.5 e$^2$fm$^4$ while the sum over 50 states is 875.9 e$^2$fm$^4$. Thus, 91\% of the strength goes to the lowest state. This is typical of what happens for all the other nuclei. Another striking feature is that the strength to the lowest 2$^+$ state and SUM(50) is consistently larger for FPD6 than for GX1A. A quantitative comparison is made in Table 18 with \% deviations ranging from 9.25\% to 19.2\% As mentioned before the single particle splittings are larger with GX1A than with FPD6 and this leads to reduced collectivity with GX1A. We note that although there is always an uncertainty of what effective charges should be used in calculating B(E2)'s. One can bring the results of B(E2) to the lowest 2$^+$ state SUM(50) with the 2 interactions closer together by making the effective charges for GX1A a bit larger than for FPD6. \\

The cumulative SUM(n) for $^{46}$Ti and $^{48}$Cr as well as the energy weighted sums are shown in Figs 2 and 3. They are smooth and do not show too much structure. However, if we look at the B(E2)'s to the next few 2$^+$ states there are discrepancies that cannot be eliminated by adjusting the effective charges. For example compare the B(E2)'s to the second and third 2$^+$ sates. In $^{46}$Ti with FPD6 they are 46.03 and 1.51 respectively while with GX1A the values are 4.03 and 43.56. Thus FPD6 predicts a strong transition to the second 2$^+$ state and a weak one to the 3rd 2$^+$ state, but GX1A gives the opposite. A similar story in $^{48}$Cr where the FPD6 numbers are 52.11 and 14.96 while the GXiA numbers are 3.09 and 75.41. On the other hand there is qualitative agreement with the 2 interactions with regards to $^{48}$Ti and $^{50}$Ti. In the former the second 2$^+$ state is the most strongly excited (108.4, 99.13), while in $^{50}$Ti it is the 3rd one (57.08, 56.93). It should be noted that the rotational model does not give much guidance for interband B(E2) transitions. If there were a simple formula for a transition from the J=0+ yrast band to the 2$^+$ state of the gamma band, Bohr and Mottelson [4] would surely have shown it. Our shell model results suggest that more experimental work should be done in this regard, just as has been done by Clement et al. [9] as shown in a previous section.

In Table 18 we show the percent deviation in the summed strength and the energy weighted strength between the 2 interactions.

\begin{table}[H]
\centering
\captionof{table}{Percent deviation $\frac{(FPD6-GX1A)}{(FPD6+GX1A)*\frac{1}{2}}$}
\begin{tabular}{|l|l|l|}
\hline
     & SUM(50) & EWS(50) \\ \hline
$^{44}$Ti & 19.03\% & 3.58\%  \\ \hline
$^{46}$Ti & 17.68\% & 4.80\%  \\ \hline
$^{48}$Ti & 11.78\% & 10.02\% \\ \hline
$^{50}$Ti & 9.25\%  & 12.46\% \\ \hline
$^{48}$Cr & 18.32\% & 10.45\% \\ \hline
\end{tabular}
\end{table}

\begin{center}

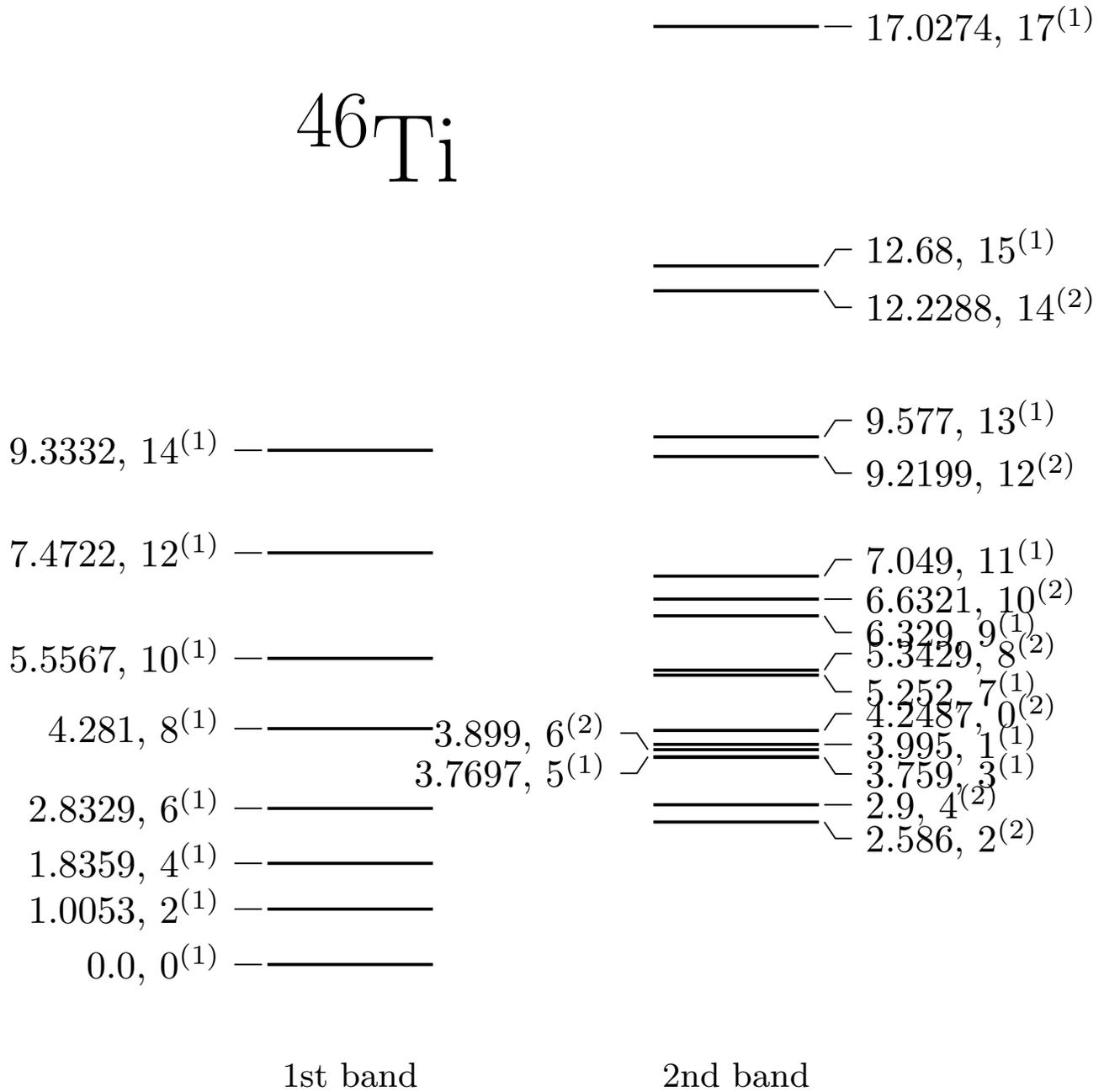
\begin{figure}[h]
\caption*{\huge{\textbf{Figure 1: $^{\textbf{46}}$Ti Energy Levels using GX1A interaction}}}
\centerline{
  \resizebox{18cm}{!}{ 
    \begin{tikzpicture}[
      scale=0.5, 
      level/.style={thick},
      virtual/.style={thick,densely dashed},
      trans/.style={thick,<->,shorten >=2pt,shorten <=2pt,>=stealth},
      classical/.style={thin,double,<->,shorten >=4pt,shorten <=4pt,>=stealth},
      label/.style = { font=\small}
    ]
    \node at (2, 15){\Huge \iso{46}{Ti}};
    \lev{0.0}{\jpi{0}{(1)}}{0};
    \lev{1.0053}{\jpi{2}{(1)}}{0};
    \lev{1.8359}{\jpi{4}{(1)}}{0};
    \lev{2.8329}{\jpi{6}{(1)}}{0};
    \lev{4.281}{\jpi{8}{(1)}}{0};
    \lev{5.5567}{\jpi{10}{(1)}}{0};
    \lev{7.4722}{\jpi{12}{(1)}}{0};
    \lev{9.3332}{\jpi{14}{(1)}}{0};
    \levS{4.2487}{\jpi{0}{(2)}}{\len+\spa}{\off};
    \levS{3.995}{\jpi{1}{(1)}}{\len+\spa}{0};
    \levS{2.586}{\jpi{2}{(2)}}{\len+\spa}{-\off};
    \levS{3.759}{\jpi{3}{(1)}}{\len+\spa}{-\off};
    \levS{2.9}{\jpi{4}{(2)}}{\len+\spa}{0};
    \levL{3.7697}{\jpi{5}{(1)}}{\len+\spa}{-\off};
    \levL{3.899}{\jpi{6}{(2)}}{\len+\spa}{\off};
    \levS{5.252}{\jpi{7}{(1)}}{\len+\spa}{-\off};
     \levS{5.3429}{\jpi{8}{(2)}}{\len+\spa}{\off};
     \levS{6.329}{\jpi{9}{(1)}}{\len+\spa}{-\off};
     \levS{6.6321}{\jpi{10}{(2)}}{\len+\spa}{0};
     \levS{7.049}{\jpi{11}{(1)}}{\len+\spa}{\off};
     \levS{9.2199}{\jpi{12}{(2)}}{\len+\spa}{-\off};
     \levS{9.577}{\jpi{13}{(1)}}{\len+\spa}{\off};
     \levS{12.2288}{\jpi{14}{(2)}}{\len+\spa}{-\off};
      \levS{12.68}{\jpi{15}{(1)}}{\len+\spa}{\off};
     \levS{17.0274}{\jpi{17}{(1)}}{\len+\spa}{0};
    \node[label] at (\len/2, -2) {1st band};
    \node[label] at (\len/2+\len+\spa, -2) {2nd band};
    \end{tikzpicture}
  }
}
\end{figure}

\clearpage

\caption*{Figure 2: SUM(n) vs n for $^{46}$Ti and $^{48}$Cr using FPD6 interaction}
\includegraphics[width=12cm,height=4.8cm]{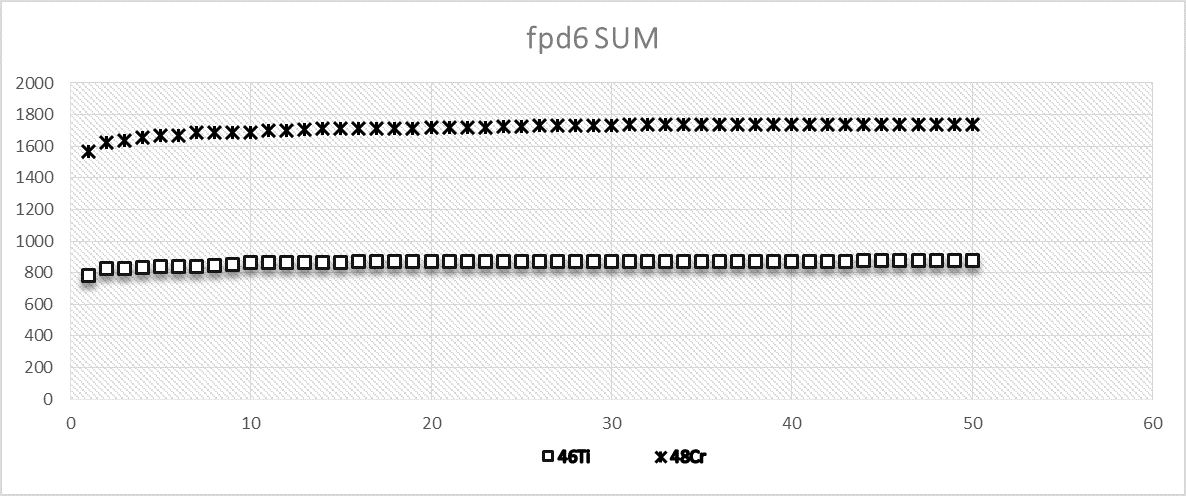}

\caption*{Figure 3: SUM(n) vs n for $^{46}$Ti and $^{48}$Cr using GX1A interaction}
\includegraphics[width=12cm,height=4.8cm]{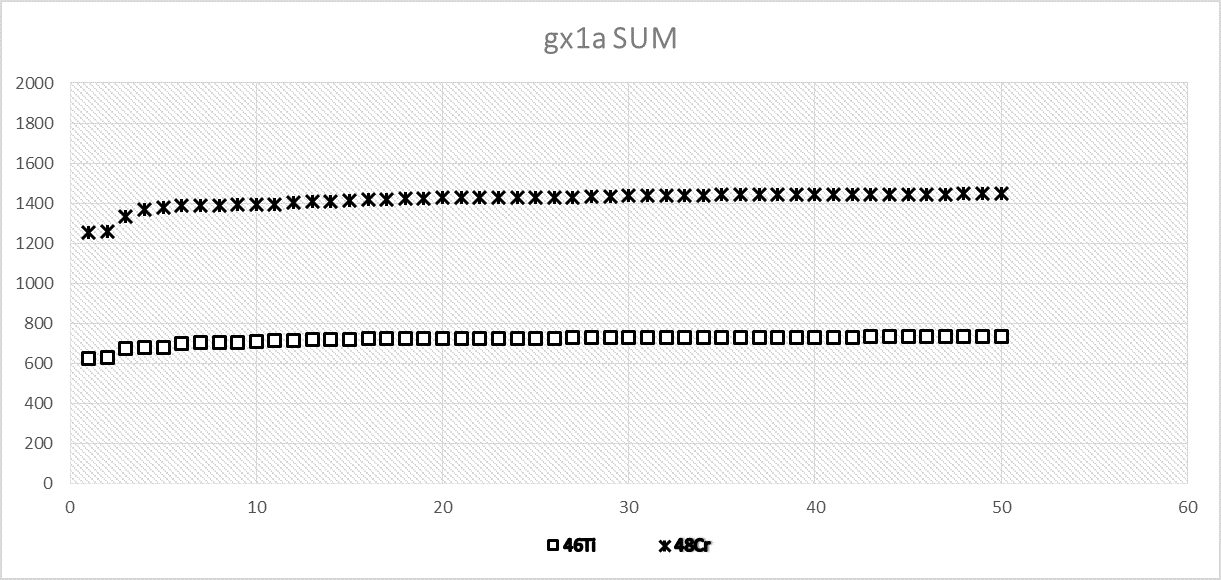}

\end{center}
In Figures 2 and 3 we show the cumulative summed strength for $^{46}$Ti and $^{48}$Cr up to 50 states. The curves show a rise at low excitation energies but they quickly flatten out which indicates that there is not much strength left. We must modify this statement by
noting that at much higher excitation energies there is a new "Giant quadrupole
strength" which our model space cannot describe. This involves excitations through two
major shells. This is discussed in many places including Bohr and Mottelson [4]. Our
model spaces have only one major shell.

\section{B(E2)s from the lowest 2$^+$ state to several 2$^+$ states}

In Tables 19 and 20 we show B(E2)'s from the 2$_1$ state to various other 2$^+$ states.
We see there is considerable fragmentation. The  2$_1$ is not really a transition but can be related to
the quadrupole moment. When we include 2$_1$ (second last column) we get a behavior inline with what we had for other result, with FPD6  yielding larger results than GX1A.
When we remove 2$_1$  (last column) for the  most part we get the opposite-GX1A bigger than FPD6. The exception is the semi-magic nucleus $^{50}$Ti. The behavior is much simpler when 2$_1$ is included in the sum.

\begin{table}[H]
\captionof{table}{B(E2)s from the 2$_1$ state to various 2$^+$ states with FPD6 interaction in
e$^{2}$fm$^{4}$}
\begin{tabular}{|l|l|l|l|l|l|l|l|}
\hline
     & 2$_1$  & 2$_2$ & 2$_3$  & 2$_4$ & 2$_5$ & SUM(50) & SUM(50) - 2$_1$ \\ \hline
$^{44}$Ti & 162.8 & 62.0 & 0.2   & 0.8  & 0.2  & 233.4   & 70.6           \\ \hline
$^{46}$Ti & 193.7 & 12.9 & 29.4  & 20.2 & 0.0  & 264.7   & 71.0           \\ \hline
$^{48}$Ti & 124.0 & 20.2 & 62.3  & 11.2 & 0.0  & 231.4   & 107.4          \\ \hline
$^{50}$Ti & 4.6   & 48.7 & 112.1 & 16.2 & 1.5  & 191.5   & 186.9          \\ \hline
$^{48}$Cr & 438.9 & 15.9 & 30.3  & 14.2 & 0.6  & 514.2   & 75.3           \\ \hline
\end{tabular}
\end{table}

\begin{table}[H]
\captionof{table}{B(E2)s from the 2$_1$ state to various 2$^+$ states with GX1A interaction in
e$^{2}$fm$^{4}$}
\begin{tabular}{|l|l|l|l|l|l|l|l|}
\hline
     & 2$_1$  & 2$_2$  & 2$_3$  & 2$_4$ & 2$_5$ & SUM(50) & SUM(50) - 2$_1$ \\ \hline
$^{44}$Ti & 12.6  & 152.9 & 0.1   & 0.1  & 2.4  & 178.5   & 166.0          \\ \hline
$^{46}$Ti & 64.6  & 105.7 & 19.2  & 8.5  & 5.2  & 213.2   & 148.6          \\ \hline
$^{48}$Ti & 72.8  & 43.7  & 54.2  & 0.6  & 0.7  & 195.0   & 122.2          \\ \hline
$^{50}$Ti & 14.8  & 35.3  & 104.0 & 1.2  & 4.4  & 172.2   & 157.4          \\ \hline
$^{48}$Cr & 330.2 & 56.5  & 8.6   & 19.8 & 0.5  & 428.9   & 98.7           \\ \hline
\end{tabular}
\end{table}

\section{Additional remarks}

There have been recent re-measurements of B(E2)'s by K. Arnswald et
al. {[}10{]} and they are somewhat different from those used for comparison
in ref {[}4{]}. The new (old) B(E2)'s for $^{44}$Ti and $^{48}$Cr
from 2$^{+}$ to 0$^{+}$ are respectively 205 (136) and 279 (274) e$^{2}$
fm$^{^{4}}$. There is a recent compilation of B(E2)'s from the lowest 2$^+$ state to ground by Pritychenko et al. {[}11{]}. This article also includes shell model calculations as support.

Early on, Gerace and Green {[}12{]} showed that admixtures
of highly deformed (intruder) states are important in the lower half
of the p-f shell and can lead to enhanced B(E2)'s. Hertz-Kintish et al. {[}13{]} noted that the measured ratio $\frac{B(E2, 4 \rightarrow 2)}{B(E2, 2 \rightarrow 0)} $ in $^{48}$Cr was smaller than the predictions of the shell, rotational
and vibrational models.

There has been recent work on vibrational spectra of even-even nuclei, especially
$^{92}$Pd {[}14,15,16{]}. Robinson et al. {[}17,18{]} made a comparison of $^{92}$Pd and $^{48}$Cr.

In working with the SU(3) Model of Elliott {[}19{]}, Kingan and Zamick
{[}20{]} noted that there are no non-zero B(E2)'s from the ground
S=0 (80) band to the S=1 (61) first excited band. This is because the
E2 operator has no spin dependence and therefore cannot connect S=0
to S=1. This is an extreme model which gives insight into why the
inter-band transitions are small. Indeed in $^{20}$Ne they found only 2 non zero B(E2)'s from the
J=0$^+$ T=0 ground state. The first was an intra-band transition to the lowest
2$^+$ state with a value of 427 e$^2$fm$^4$, while the second was to a T=1 (S=0) state with a
value of 12.56 e$^2$fm$^4$. The smallness of the latter can be explained by the fact that
T=0 to T=0 B(E2)'s are proportional to (e$_p$ +e$_n$)$^2$ whereas T=0 to T=1 are
proportional to (e$_p$ - e$_n$)$^2$, where the e's are effective charges.

\section{Acknowldgements}

We thank Mihai Horoi for helpful comments. We thank Stephan Frauendorf and Terry Fortune for their
comments. Levering Wolfe was supported by the Aresty Research Assistant Program and by a Richard J. Plano Internship.


\begin{thebibliography}{10}
\bibitem[1]{key-7} B. F. Bayman, J. D. McCullen, and Larry Zamick Phys. Rev. Lett. 11, 215 (1963)

\bibitem[2]{key-6} J. D. McCullen, B. F. Bayman, and Larry Zamick Phys. Rev. 134, B515 (1964)

\bibitem[3]{key-5}R. Sherr, B. F. Bayman, E. Rost, M. E. Rickey, and C. G. Hoot Phys. Rev. 139, B1272 (1965)

\bibitem[4]{key-2}A Bohr and B.R. Mottelson, Nuclear Structre II:
Nuclear Deformation,World Scientific, Singapore (1975).


\bibitem[5]{key-7} S.J.Q . Robinson, T. Hoang, L. Zamick, Y.Y. Sharon
and A.Escuderos Phys. Rev. C89, 014316 (2014).

\bibitem[6]{key-5} The Shell Model Code NUSHELLX@MSU, B.A. Brown
and W.D.M. Rae, http://www.sciencedirect.com/science/article/pii/S0090375214004748

\bibitem[7]{key-3-2}M. Honma, T. Otsuka, B. A. Brown, and T. Mizusaki,
Phys. Rev. C65 ,061301 (R),2002; Phys. Rev. C 69, 034335 (2004).

\bibitem[8]{key-1} W.A. Richter, M.G. Vadermerwe, R.E. Julius and
B.A. Brown,Nuc; Phys.A 704, 134(2002)


\bibitem[9]{key-4} E. Cl\'ement et al.Phys. Rev. C 94, 054326 (2016)

\bibitem[10]{key-1}K. Arnswald et al. , Phys.Lett.B772(2017)599

\bibitem[11]{key-2} B. Pritychenko, M. Birch, M. Horoi and R. Singh, Nuclear Data Sheets 120 (2014) 112

\bibitem[12]{key-1}W.J. Gerace and A.M. Green, Nucl. Phys. A113,641(1958)

\bibitem[13]{key-3}D. Hertz-Kintish, L.Zamick and S.J.Q.Robinson, Phys.
Rev. C90,030307 (2014)

\bibitem[14]{key-11} B. Cederwall et al., Nature 469:68 (2011)

\bibitem[15]{key-12} Chong Qi, J. Blomqvist, T. B\"ack, B. Cederwall, A. Johnson, R. J. Liotta, R. Wyss Phys. Rev. C84: 021301, (2011)

\bibitem[16]{key-6} S.G. Frauendorf and A.O. Macchiavelli, Progress in Nuclear and Particle Physics 78, (2014) 24

\bibitem[17]{key-14} S.J.Q. Robinson, T. Hoang, L. Zamick, A. Escuderos, and Y.Y. Sharon, Phys. Rev. C 89, 014316 (2014)

\bibitem[18]{key-15} S.J.Q. Robinson, M. Harper and L. Zamick, nucl-th arXiv:411.1390 (2016)

\bibitem[19]{key-1}J.P. Elliott, Proc. R. Soc. London, Ser .A 245,128
(1958)

\bibitem[20]{key-2}A.Kingan and L.Zamick, Int. Jour. Modern Physics
E,27 (2018) 1850056



\end{thebibliography}
\end{document}